\documentclass[reprint,two-column,double-spaced,groupedaddress,superscriptaddress,pra]{revtex4-2}

\usepackage[normalem]{ulem}

\usepackage{mathrsfs}
\usepackage{enumerate}
\usepackage{graphicx}
\usepackage{physics}

\usepackage[colorlinks]{hyperref}
\usepackage{tensor}
\usepackage{leftidx}
\usepackage{braket}
\usepackage{amssymb} 

\begin{document}
\title{Photonic non-Markovianity identification \\by quantum process capabilities of non-CP processes}\date{\today}

\author{Chan Hsu}
\affiliation{Department of Engineering Science, National Cheng Kung University, Tainan 701401, Taiwan}
\affiliation{Center for Quantum Frontiers of Research $\&$ Technology, National Cheng Kung University, Tainan 701401, Taiwan}

\author{Yu-Chien Kao}
\affiliation{Department of Engineering Science, National Cheng Kung University, Tainan 701401, Taiwan}
\affiliation{Center for Quantum Frontiers of Research $\&$ Technology, National Cheng Kung University, Tainan 701401, Taiwan}

\author{Hong-Bin Chen}
\affiliation{Department of Engineering Science, National Cheng Kung University, Tainan 701401, Taiwan}
\affiliation{Center for Quantum Frontiers of Research $\&$ Technology, National Cheng Kung University, Tainan 701401, Taiwan}
\affiliation{Physics Division, National Center for Theoretical Sciences, Taipei 10617, Taiwan}

\author{Shih-Hsuan Chen}
\affiliation{Department of Engineering Science, National Cheng Kung University, Tainan 701401, Taiwan}
\affiliation{Center for Quantum Frontiers of Research $\&$ Technology, National Cheng Kung University, Tainan 701401, Taiwan}

\author{Che-Ming Li}
\email{cmli@mail.ncku.edu.tw}
\affiliation{Department of Engineering Science, National Cheng Kung University, Tainan 701401, Taiwan}
\affiliation{Center for Quantum Frontiers of Research $\&$ Technology, National Cheng Kung University, Tainan 701401, Taiwan}
\affiliation{Physics Division, National Center for Theoretical Sciences, Taipei 10617, Taiwan}
\affiliation{Center for Quantum Science and Technology, Hsinchu 300044, Taiwan}

\begin{abstract}
A Markovian quantum process can be arbitrarily divided into two or more legitimate completely-positive (CP) subprocesses. When at least one non-CP process exists among the divided processes, the dynamics is considered non-Markovian. However, how to utilize minimum experimental efforts, without examining all process input states and using entanglement resources, to identify or measure non-Markovianity is still being determined. Herein, we propose a method to quantify non-CP processes for identifying and measuring non-Markovianity without the burden of state optimization and entanglement. This relies on the non-CP processes as new quantum process capabilities and can be systematically implemented by quantum process tomography. We additionally provide an approach for witnessing non-Markovianity by analyzing at least four system states without process tomography. We faithfully demonstrate that our method can be explicitly implemented using all-optical setups and applied to identify the non-Markovianity of single-photon and two-photon dynamics in birefringent crystals. Our results also can be used to explore non-Markovianity in other dynamical systems where process or state tomography is implementable.
\end{abstract}


\maketitle

\section{Introduction}\label{sec:introduction}
In real-world scenarios, open quantum systems are inevitably affected by their environments. The classification of the dynamical behaviors of open quantum systems based on the observed physical phenomena and the type of dynamical mapping in mathematical form has been extensively studied \cite{tool1,tool2,Kraus,BandP,chen2009,zhang2012,yin2012,chruscinski2014,chen2015hierarchy,chen2015using,xiong2015,breuer2016colloquium,urrego2018,HongBin2019}. Markovian and non-Markovian dynamics are seminal paradigms illustrating different dynamical behaviors \cite{breuer2009,laine2010,rivas2010,luo2012,breuer2012,rivas2014,guarnieri2014quantum,breuer2016colloquium,chen2016,deVega2017,li2019non,wu2020}. 

From the perspective of the information flow caused by the system-environment interactions, Markovian dynamics leads to a monotonically decreasing trace distance between two system states \cite{breuer2009,laine2010} and may thus also lead to a monotonic loss of entanglement between the principal system and an isolated ancilla \cite{rivas2010} together with a monotonic reduction of mutual information between them \cite{luo2012}. In contrast, non-Markovian dynamics allows the backflow of information from the environment to the system and may thus lead to a  temporal revival in these quantities \cite{breuer2009,laine2010,rivas2010,luo2012}. Other qualities affected include the Fisher information \cite{lu2010quantum}, Bloch volume \cite{lorenzo2013geometrical}, Gaussian channels \cite{torre2015non}, and quantum probing of convex coefficients \cite{Lyyra22}.

Among existing criteria aimed at identifying or measuring non-Markovianity by analyzing the output states after the finite input states undergo the dynamical process, the BLP (Breuer, Laine, and Piilo) criterion \cite{breuer2009,laine2010}  highlights that the revival of the trace distance between two system states undergoing an experimental process is a signature of the non-Markovianity of the dynamics. With this fact, the BLP criterion provides a quantitative description of non-Markovianity by maximization over all possible initial state pairs. The single-photon and two-photon experiments in Refs.~\cite{liu2011,liu2013}, which were performed with birefringent crystals and an all-optical setup, used the BLP criterion \cite{breuer2009,laine2010} to characterize the photon dynamics and experimentally illustrate the control of the transition from Markovian to non-Markovian dynamics. The two-photon experiment in Ref.~\cite{liu2013} additionally investigated the realization of the nonlocal memory effect \cite{laine2012}, in which, for a composite open system composed of two subsystems and their environments, the local dynamics of each subsystem is Markovian while the global dynamics is non-Markovian.

While the success in the experimental demonstrations of the BLP criterion, for an unknown open quantum system, the maximization over all possible initial state pairs becomes experimentally tricky to implement the measure based on the BLP criterion. Equivalently, it is also tricky to choose initial state pairs to perform trials to identify the revival of the trace distance between the two system states for non-Markovianity.

From the perspective of the mathematical properties of the dynamical map, the Markovian dynamics defined in the time-local master equation \cite{lindblad1976generators, gorini1976completely} for dynamics involving time-dependent principal system and system-environment interactions satisfies completely-positive (CP) divisibility. A Markovian process can be arbitrarily divided into two or more legitimate CP subprocesses. By contrast, non-Markovian dynamics causes one or more of the intermediate processes to be non-CP and therefore does not satisfy CP-divisibility. The RHP (Rivas, Huelga, and Plenio) criterion \cite{rivas2010} enables identification and measure of non-Markovian dynamics by the CP-divisibility. However, implementing the RHP criterion experimentally is challenging due to the requirement to achieve the maximally entangled state of the principal system and the ancilla and entanglement-related operations. This meams the RHP criterion requires no optimization concerning other related resources, unlike the BLP criterion. On the other hand, from the viewpoint of a weak non-Markovian process, almost all experimentally implementable methods can only be used to detect non-P divisible instead of some P-divisible but non-CP divisible process \cite{Hall14}. The information flow condition in the BLP criterion can be considered as a slightly weaker version of the P-divisible criterion (see Eq.~(3) in Ref.~\cite{Dariusz18} and the related discussions.)

Compared to the BLP and the RHP criteria, this study introduces and utilizes new quantum process capabilities (QPCs) of non-CP processes for identifying and measuring non-Markovian dynamics without examining all initial system states nor using entangled pairs. The QPC formalism \cite{QPC} provides an experimentally implementable method of quantitatively characterizing quantum operations according to a specified process characteristic. Here, the QPCs are quantified by the robustness of non-CP processes to endure the minimum amounts of CP operations required to become CP processes. This satisfies all the conditions for sensible measures of process capabilities \cite{QPC}. With this finding, non-Markovianity can also be quantified by these QPCs of non-CP processes, where a non-Markovian process can manifest itself in non-Markovianity by the QPC of non-CP process. In contrast, a Markovian quantum process is considered without non-Markovianity. Experimentally, they can be systematically measured using quantum process tomography (QPT) \cite{NandC}. Furthermore, we show how to witness them by analyzing at least four system states without QPT.

The introduced concept and method are applied to analyze the non-Markovianity of single-photon and two-photon dynamics in birefringent crystals using all-optical setups. Following our previous study \cite{WangKH}, which systematically utilized the experimental data reported in Ref.~\cite{liu2011} to implement QPT for thoroughly describing the single-photon dynamics in terms of process matrix, we reliably reconstructed the process matrices in the present study for dynamics in the existing two-photon experiment \cite{liu2013}. Therefore, by measuring the QPCs of non-CP processes, we faithfully identify non-Markovianity in the single-photon and two-photon experimental processes \cite{laine2012,liu2013}. This also shows that QPC formalism provides practical and general utility to quantify the non-physicality of quantum maps~\cite{npQM}.

In the following, we first define non-Markovian dynamics and non-CP process in Sec.~\ref{2}. Then, we introduce the new QPCs of non-CP processes and the methods of identifying and measuring non-Markovianity. Appendix~\ref{appendix_rev} compares the implementation requirements of BLP and RHP criteria for non-Markovianity. In Sec.~\ref{3}, we detail how we implement our method to faithfully identify non-Markovianity of photon dynamics in birefringent crystals reported in Refs.~\cite{liu2011,liu2013}. Appendix~\ref{appendix_exp} shows how to use the reported experimental data \cite{liu2011,liu2013} to construct exact process matrices for our non-Markovianity identification. While the case of the pure dephasing process has been investigated in Refs.~\cite{liu2011,liu2013}, and our theory is not limited to such a case, here we mainly show how our results are experimentally applicable to a photonics system. As will be shown, our illustrations of the proposed concept and methods can be reliably and faithfully delivered with the photonics experimental data provided in Refs.~\cite{liu2011,liu2013} and without performing a new experiment to show the practical feasibility of our results. In Sec.~\ref{4}, we present a witness to detect non-Markovianity and demonstrate which in photon dynamics in the experiments of Refs.~\cite{liu2011,liu2013}. Insights and the outlook of our work have been summarized in Sec.~\ref{5}.


\begin{figure}[t]
\includegraphics[width=0.35\textwidth]{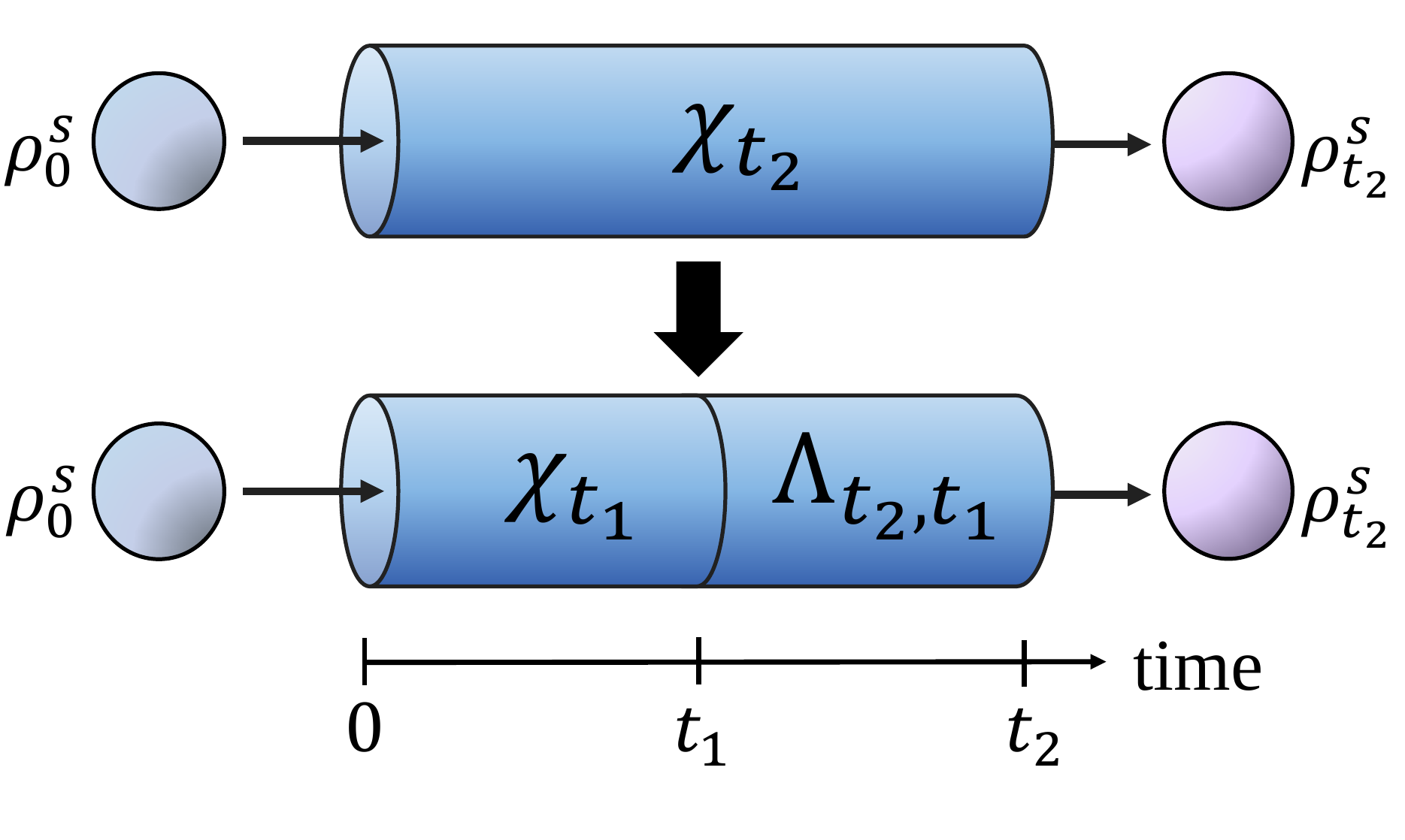}
\caption{Illustration of the CP-divisibility concept in Markovian dynamics with CP $\Lambda_{t_2,t_1}$ and the non-CP processes for non-Markovian dynamics with non-CP $\Lambda_{t_2,t_1}$.}
\label{F-Determining}
\end{figure}

\section{non-Markovianity and quantum process capabilities}\label{2}
\subsection{Non-Markovian dynamics and non-CP processes}\label{3.1.1}

The time-local master equation with a time-dependent generator $\mathcal{K}_{t}$ \cite{lindblad1976generators, gorini1976completely}, has the form
\begin{eqnarray}  \label{timelocal}
\begin{aligned}
&\frac{d}{dt} \rho^{S}_t = \mathcal{K}_{t} [\rho^{S}_t] = -i [H(t), \rho^{S}_t] \\
& +\!\!\sum_{k}\!\gamma_{k}(t)\!\left[ L_{k}(t) \rho^{S}_t L^{\dagger}_{k}(t)\!-\!\frac{1}{2} \lbrace L^{\dagger}_{k}(t)L_{k}(t), \rho^{S}_t \rbrace  \right],
\end{aligned}
\end{eqnarray}
where the Hamiltonian $H(t)$, Lindblad operators $L_{k}(t)$, and relaxation rates $\gamma_{k}(t)$ may be time-dependent. A dynamics $\mathcal{E}_t=\mathcal{T}\exp[\int_0^t\mathcal{K}_\tau d\tau]$ generated by the time-local master equation~(\ref{timelocal}) is defined to be \emph{Markovian} if it is CP-divisible, i.e., all of the intermediate mapping $\Lambda_{t_2,t_1}$ satisfying
\begin{equation}\label{CP-divisibility}
\mathcal{E}_{t_2}=\Lambda_{t_2,t_1}\circ\mathcal{E}_{t_1}
\end{equation}
being CP $\forall$ $0\leq t_1\leq t_2$. Consequently, a dynamics $\mathcal{E}_t$ is defined to be \emph{non-Markovian} if one finds that $\Lambda_{t_2,t_1}$ is non-CP for certain $t_2$ and $t_1$. As represented this form in process matrices \cite{NandC}, that is,
\begin{equation}\label{divisibility}
\chi_{t_{2}}=\Lambda_{t_{2},t_{1}} \circ \chi_{t_{1}},
\end{equation}
where the original process $\chi_{t_{2}}$ from time point 0 to $t_{2}$ is divided into a subprocess $\chi_{t_{1}}$ from time point 0 to $t_{1}$ for arbitrary time $t_{1}$ and an intermediate process from time point $t_{1}$ to $t_{2}$ described by the process matrix, denoted as $\Lambda_{t_{2},t_{1}}$. See Fig.~\ref{F-Determining}.

One can use QPT to characterize $\chi_{t_{2}}$ and $\chi_{t_{1}}$ and calculate the intermediate process $\Lambda_{t_{2},t_{1}}$ by $\chi_{t_{2}}$ and the inverse matrix of $\chi_{t_{1}}$ as follows:
\begin{equation}\label{Determining}
\Lambda_{t_{2},t_{1}}= \chi_{t_{2}} \chi_{t_{1}}^{-1}.
\end{equation}
If the dynamical process under consideration is Markovian, the original process $\chi_{t_{2}}$ satisfies the CP-divisibility, where $\Lambda_{t_{2},t_{1}}$ is a CP process and all of its eigenvalues are non-negative \cite{breuer2016colloquium}. However, if even one of the eigenvalues is negative, then $\Lambda_{t_{2},t_{1}}$ is a non-CP process, and hence $\chi_{t_{2}}$ is non-Markovian dynamics. It is noted that this identification method is qualitative rather than quantitative in nature. The degree of non-CP process and non-Markovianity can be further quantified using the method introduced below in Sec. \ref{3.3.2}.

\subsection{Quantum process capabilities of non-CP processes for non-Markovianity}\label{3.3.2}

\subsubsection{Non-CP processes as quantum process capabilities}\label{3.3.2.1}

In theory, a non-CP intermediate process $\Lambda_{t_{2},t_{1}}$ can become a CP process by mixing with a CP process $\chi'$. Thus, $\Lambda_{t_{2},t_{1}}$ can be quantitatively characterized by the minimum amount of $\chi'$, denoted as $\beta$, called robustness, which must be added to it such that it becomes CP process $\chi_{\text{CP}}$. That is,
\begin{equation}\label{beta}
\frac{\Lambda_{t_{2},t_{1}}+\beta\chi'}{1+\beta}=\chi_{\text{CP}}.
\end{equation}

For a given $\Lambda_{t_{2},t_{1}}$ (\ref{Determining}), $\beta$ can be determined by solving the following optimization equation via semidefinite programming (SDP) with MATLAB \cite{sdp1,sdp2}:
\begin{equation}\label{B_eq}
\beta \!=\!\min_{\tilde{\chi}_{\text{CP}}}[\,\text{tr}(\,\tilde{\chi}_{\text{CP}})\,-1]\,,
\end{equation}
under the constraints:
\begin{equation}\label{Bconst1}
\tilde{\chi}_{\text{CP}} \geq 0,
\end{equation}
\begin{equation}\label{Bconst2}
\tilde{\chi}_{\text{CP}}-\chi_{t_{2}} \chi_{t_{1}}^{-1} \geq 0,
\end{equation}
\begin{equation}\label{Bconst3}
\text{tr} (\,\tilde{\chi}_{\text{CP}})\, \geq 1.
\end{equation}
The first constraint guarantees that $\tilde{\chi}_{\text{CP}}$ is CP, while the second constraint ensures that $\chi'$ is also CP. The third constraint guarantees that $\beta \geq 0$. For instance, when $\Lambda_{t_{2},t_{1}}$ is a non-CP process,  $\beta$ is greater than zero and equal to the sum of the absolute values of all the negative eigenvalues of $\Lambda_{t_{2},t_{1}}$. Conversely, if $\Lambda_{t_{2},t_{1}}$ is already a CP process, no additional operations $\chi'$are required, and $\beta$ is equal to zero. In the following, we demonstrate that the non-CP process can be regarded as a type of QPC through its robustness $\beta$ [Eq.~(\ref{beta})]. 

From the QPC theory \cite{QPC}, a non-CP process can be regarded as a capable process showing the non-CP characteristic (denoted as $\chi_{\mathcal{C}}$). Conversely, a process without non-CP characteristics can be regarded incapable (denoted as $\chi_{\mathcal{I}}$). The robustness $\beta$ satisfies the following properties for a sensible QPC measure \cite{QPC}: $C(\Lambda_{t_{2},t_{1}})$, to faithfully quantify the non-CP characteristic:\\

\noindent(MP1) Faithfulness: $C(\Lambda_{t_{2},t_{1}})=0$ if and only if $\Lambda_{t_{2},t_{1}}$ is incapable of manifesting the non-CP characteristic.\\

\noindent(MP2) Monotonicity: $C(\Lambda_{t_{2},t_{1}}\circ \chi_{\text{CP}}) \leq C(\Lambda_{t_{2},t_{1}})$. The degree of non-CP characteristic of $\Lambda_{t_{2},t_{1}}$ does not increase when applying additional incapable processes, $\chi_{\text{CP}}$.\\

\noindent(MP3) Convexity: $C(\sum_{n}p_{n}\Lambda_{t_{2},t_{1}}\circ \chi_{\text{CP}n}) \leq \sum_{n}p_{n}C(\Lambda_{t_{2},t_{1}}\circ \chi_{\text{CP}n})$. Mixing processes does not increase the strength of non-CP characteristic.\\

The proofs of the robustness satisfying (MP1), (MP2) and (MP3) are provided as follows:\\

\noindent(P1) Proof of (MP1). The robustness satisfies (MP1) directly according to the definition of $\beta$. That is, when $\Lambda_{t_{2},t_{1}}$ is a CP process, (i.e., an incapable process), the value of $\beta$ is zero.\\

\noindent(P2) Proof of (MP2). The process of $\Lambda_{t_{2},t_{1}}$ incorporated with incapable processes $\chi_{\mathcal{I}n}$ can be represented in the form shown in Eq.~(\ref{beta}) as follows:
\begin{eqnarray}
\begin{aligned}
\label{MP2}
\Lambda_{t_{2},t_{1}} \circ \chi_{\text{CP}} & =[(1+\beta)\chi_{\text{CP}}-\beta \chi'] \circ \chi_{\text{CP}} \\
& =(1+\beta)\chi_{\text{CP}} \circ \chi_{\text{CP}}-\beta \chi' \circ \chi_{\text{CP}} \\
& =(1+\beta')\chi_{\text{CP}}-\beta' \chi'.
\end{aligned}
\end{eqnarray}
Since $\chi_{\text{CP}}$ is an incapable process, $\chi_{\text{CP}} \circ \chi_{\text{CP}}$ must also be an incapable process. Thus, the optimal $\beta'$ value of $\Lambda_{t_{2},t_{1}} \circ \chi_{\text{CP}}$ must be equal to or less than the optimal $\beta$ value of $\Lambda_{t_{2},t_{1}}$, i.e., $\beta' \leq \beta$.\\

\noindent(P3) Proof of (MP3). Let $\sum_{n}p_{n}\Lambda_{t_{2},t_{1}}\circ \chi_{\text{CP}n}$ be represented as\\
\begin{eqnarray}
\begin{aligned}
\label{MP3}
\sum_{n}p_{n}\Lambda_{t_{2},t_{1}}\circ \chi_{\text{CP}n} =(1+b')\chi_{\text{CP}'}-b' \chi'',
\end{aligned}
\end{eqnarray}
where
\begin{eqnarray}
\begin{aligned}
\label{MP3-1}
\chi_{\text{CP}'} & = \frac{\sum_{n}p_{n}(1+\beta_{n})\chi_{\text{CP}n}}{1+b'}, \\
\chi'' & = \frac{\sum_{n}p_{n}\beta_{n}\chi'_{n}}{b'}, \\
b' & =\sum_{n}p_{n}\beta_{n}=\sum_{n}p_{n}C(\Lambda_{t_{2},t_{1}}\circ \chi_{\text{CP}n}),
\end{aligned}
\end{eqnarray}
and $\chi'_{n}$ is the CP process for each $\Lambda_{t_{2},t_{1}}\circ \chi_{\text{CP}n}$. Since $\beta = C(\sum_{n}p_{n}\Lambda_{t_{2},t_{1}}\circ \chi_{\text{CP}n})$ is the optimal $b$, then $\beta \leq b'$.

\subsubsection{Non-Markovianity quantified by \\quantum process capabilities of non-CP processes }

From the definitions given in Secs.~\ref{3.1.1} and \ref{3.3.2.1}, it is clear that a non-Markovian dynamics $\chi_{t_2}$ has \emph{at least one} non-CP process, $\Lambda_{t_{2},t_{1}}$, which exists among the divided processes and possesses robustness $\beta>0$ for a certain $t_1$. Whereas a Markovian quantum process satisfies the CP divisibility, its $\Lambda_{t_{2},t_{1}}$ is CP with $\beta=0$, $\forall$ $0\leq t_1\leq t_2$. Therefore, according to the QPC theory~(Sec.~\ref{3.3.2.1}), the former can show non-Markovianity, and the latter cannot, as the sensible QPC measure quantifies. See Sec.~\ref{3.3.3} for detailed discussions of how to use QPCs of non-CP processes to detect non-Markovianity. Specifically, the QPC theory is not directly associated with the non-Markovianity itself; instead, it circuitously via the non-CP characteristic of the intermediate process in the composition law Eq.~(\ref{CP-divisibility}).\\

\subsection{Identifying and measuring non-Markovianity}\label{3.3.3}

\subsubsection{Non-Markovianity idenitification}

With the non-Markovianity quantified by the robustness $\beta$, one can therefore identify a process $\chi_{t_2}$ as non-Markovian if the robustness of the intermediate process $\Lambda_{t_{2},t_{1}}$ in Eq.~(\ref{Determining}) is positive for a chosen time $t_1$, namely,
\begin{equation}
 \beta(\Lambda_{t_{2},t_1})>0.\label{identification} 
\end{equation}
Experimentally, as described in Eq.~(\ref{Determining}), performing such identification of non-Markovianity requires QPT to obtain the process matrices $\chi_{t_{2}}$ and $\chi_{t_{1}}$ and a calculation of the intermediate process $\Lambda_{t_{2},t_{1}}$ by $\chi_{t_{2}}$ and the inverse matrix of $\chi_{t_{1}}$.

\subsubsection{Non-Markovianity measure}

A non-Markovian quantum dynamics should have at least one non-CP intermediate process $\Lambda_{t_{2},t_1}$ for a specific $t_1$. Each such $\Lambda_{t_{2},t_1}$ can be quantified by the sensible measure $\beta$. For Markovian processes, we have $\beta(\Lambda_{t_{2},t_1})=0$. Therefore it is objective to define a measure of the total dynamics $\chi_{t_{2}}$ by summing all the robustness of $\Lambda_{t_{2},t_1}$ for $0\leq t_1\leq t_2$. The overall non-Markovianity of the process $\chi_{t_2}$ is then quantitatively described by
\begin{equation}\label{betanonMK}
\mathcal{N}_{\rm \beta} =\int^{t_2}_{0} \beta(\Lambda_{t_{2},t}) dt.
\end{equation}

\subsubsection{Comparison}

Compared with the BLP criterion-based non-Markovianity identification and measure, the experimental requirement for testing Eq.~(\ref{identification}) and measuring $\mathcal{N}_{\rm \beta}$ is deterministic by QPT without optimizing the input states of the process $\chi_{t_{2}}$. Moreover, our method requires no entanglement and the related entanglement manipulation needed by the non-Markovianity identification and measure using the RHP criterion. See Appendix~\ref{appendix_rev} for a review of these two methods. 

\section{Non-Markovianity identification of photon dynamics}\label{3}

To show the concept and method presented in Sec.~\ref{2} being experimentally feasible without losing generality, we specifically demonstrate how to identify non-Markovian photon dynamics in birefringent crystals using the idea underlying Eq.~(\ref{identification}). We will first describe how we reliabily get the required process matrices of photon dynamics from the existing experiments \cite{liu2011,liu2013}. Then the results of non-Markovianity identification will be given afterward. With these demonstrations as a basis, experimentally, it is feasible to implement the non-Markovianity measure [Eq.~(\ref{betanonMK})] by summing a sufficient number of the measured robustness $\beta(\Lambda_{t_{2},t})$.

\subsection{Process matrices of photon dynamics}

\subsubsection{Single-photon experiment}

\begin{figure*}[t]
\includegraphics[width=18cm]{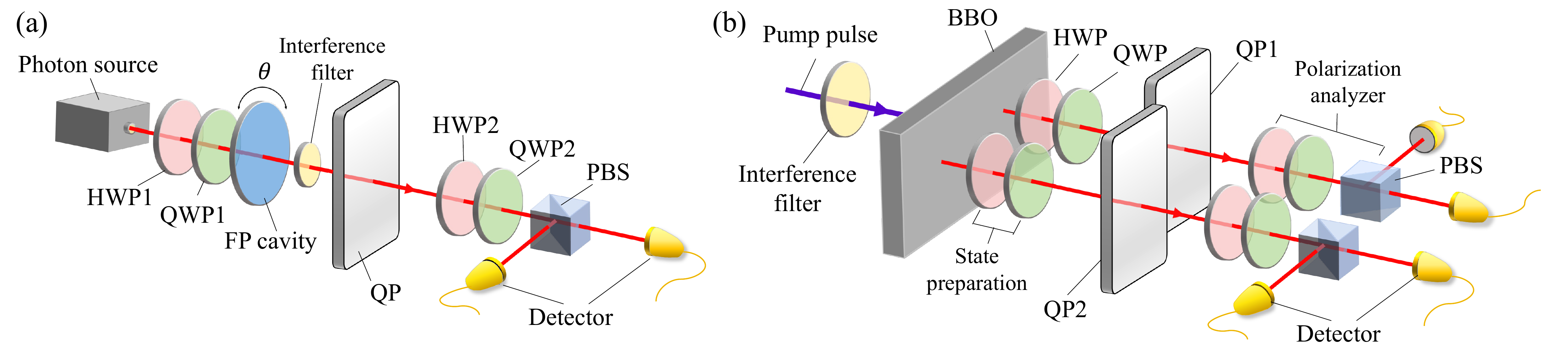}
\caption{Schematic illustration of the experimental setups used to identify photonic non-Markovianity. (a) single-photon dynamics in a birefringent crystal of quartz plate (QP) \cite{liu2011}. (b) two-photon dynamics in birefringent crystals of quartz plates (QP1,QP2)~\cite{liu2013}.}
\label{exp}
\end{figure*}

In Fig.~\ref{exp}(a), the polarization degree of freedom of the photon is regarded as the principal system and is prepared with an arbitrary state $\ket{\lambda}$ using a half-wave plate (HWP1) and quarter-wave plate (QWP1). The tomography of the polarization states was measured by a pair of polarization analyzers; each consisting of a QWP (QWP2), a HWP (HWP2), and a polarizing beam splitter (PBS), and is then detected by two single-photon detectors. Moreover, the Fabry Pérot (FP) cavity is used to optionally prepare the frequency state $\ket{\Omega^{\theta}}=\int d \omega f^{\theta}(\omega)\ket{\omega}$ by adjusting its angle $\theta$, which is treated as the environment of the principal system. The function $f^{\theta}(\omega)$ represents the amplitude of the mode with frequency $\omega$, where the frequency distribution $\left| f^{\theta}(\omega) \right| ^2$ satisfies the normalization condition $\int d \omega \left| f^{\theta}(\omega) \right| ^2=1$. These two degrees of freedom are coupled in a quartz plate. The time evolution of the principal system and environment in the quartz plate can thus be described by the following unitary transformation $U(t)$:
\begin{equation}\label{1.1}
U(t) \ket{\lambda} \otimes \ket{\Omega^{\theta}} = \int d \omega f^{\theta}(\omega) e^{i \omega n_{\lambda} t} \ket{\lambda} \otimes \ket{\omega},
\end{equation}
where $\lambda=H, V$ denotes horizontal or vertical polarization, respectively; and $n_{\lambda}$ is the refraction index of the corresponding polarized light $\lambda$ in the quartz plate.

When considering the dynamical changes of the polarization states over time, let us assume that the principal system and its environment are initially prepared in the states $\rho^{S}_{0}$ and $\rho^{E}_{0}=\ket{\Omega^{\theta}}\!\!\bra{\Omega^{\theta}}$, respectively. We use the quantum operations formalism \cite{Kraus,NandC,breuer2002} to construct a dynamical map relating $\rho^{S}_{0}$ to $\rho^{S}_{t}$ by
\begin{equation}\label{1.3}
\rho^{S}_{t}=\chi^{\theta}_{t}(\rho^{S}_{0})=\sum_{m=1}^{4}\sum_{n=1}^{4}\chi^{\theta}_{t,mn}M_{m}\rho^S_0M_{n}^{\dagger},
\end{equation}
where $M_{1}=I$, $M_{2}=X$, $M_{3}=-iY$, $M_{4}=Z$, and $I$ and $X$, $Y$, $Z$ denote the identity operator and the Pauli matrices, respectively. Here $\chi^{\theta}_{t}$ composed of the coefficients $\chi^{\theta}_{t,mn}$ is the process matrix for the dynamical map of the polarization state:
\begin{equation}\label{1.X}
\chi^{\theta}_{t}=\frac{1}{4}
\left[
\begin{matrix}
2+\kappa^{\theta}(t)+\kappa^{\theta*}(t)  & 0  & 0  & \kappa^{\theta}(t)-\kappa^{\theta*}(t) \\
0                                                             & 0  & 0  & 0\\
0                                                             & 0  & 0  & 0\\
\kappa^{\theta*}(t)-\kappa^{\theta}(t)      & 0  & 0  & 2-\kappa^{\theta}(t)-\kappa^{\theta*}(t)\\
\end{matrix}
\right].
\end{equation}
The decoherence factor $\kappa^{\theta}(t)$ in Eq. (\ref{1.X}) is given by the corresponding frequency distribution at a specific angle $\theta$ of the FP cavity, i.e.,
\begin{equation}\label{1.k}
\kappa^{\theta}(t) = \int d \omega \left| f^{\theta}(\omega) \right| ^2 e^{i \omega \Delta n t},
\end{equation}
where $\Delta n={n}_{V}-{n}_{H}$ depends on the particular specification of the quartz plate.

\subsubsection{Two-photon experiment}\label{3.2.1}

The identification of non-Markovian dynamics in two-photon scenarios was investigated using the experimental system implemented in Ref.~\cite{liu2013}, consisting of entangled photon pairs generated through a spontaneous parametric down-conversion (SPDC) process in a $\beta$-barium borate (BBO) crystal with an ultraviolet pump pulse. See Fig.~\ref{exp}(b). An interference filter was used to control the spectral width $\delta$ of the pump pulse and prepare the frequency state $\ket{\Omega}=\int d\omega_{1}d\omega_{2}g(\omega_{1},\omega_{2})\ket{\omega_1}\!\otimes\!\ket{\omega_2}$. Note that $g(\omega_{1},\omega_{2})$ denotes the probability amplitude of finding a state $\ket{\omega_1}\!\otimes\!\ket{\omega_2}$ of a photon pair with frequency $\omega_{1}$ in spatial mode 1 and a photon with frequency $\omega_{2}$ in spatial mode 2.

The two degrees of freedom of each photon were then locally coupled in two quartz plates (QP1, QP2). The local interaction in each quartz plate at an active time $t$ can be described by the unitary transformation $U_{i}(t)\ket{\lambda}\otimes\ket{\omega_{i}}=e^{i{\omega_{i}}n_{\lambda}t}\ket{\lambda}\otimes\ket{\omega_{i}}$, for $i=1, 2$. Note that the active times of the single-photon interactions in QP1 and QP2, respectively, are non-overlapping and consecutive. In particular, QP1 activates before QP2. Finally, the tomography of the polarization states was measured by a pair of polarization analyzers.

We assume that the principal system and its environment are initially prepared in the states $\rho^S_0$ and $\rho^{E}_0=\ket{\Omega}\!\!\bra{\Omega}$, respectively. According to the quantum operation formalism \cite{Kraus,NandC,breuer2002}, the final state of the system can be represented as
\begin{equation}\label{2.3}
\rho^{S}_{t}=\chi^{\delta}_{t}(\rho^{S}_{0})=\sum_{m=1}^{16}\sum_{n=1}^{16}\chi^{\delta}_{t,mn}E_{m}\rho^{S}_0E_{n}^{\dagger},
\end{equation}
where $E_{m(n)}=\ket{h}\!\!\bra{r}\otimes\ket{l}\!\!\bra{s}$, in which $m(n)=s\cdot 2^{0}+r\cdot 2^{1}+l\cdot 2^{2}+h\cdot 2^{3}+1$ for $h,r,l,s\in\{0,1\}$ and $\ket{0}\equiv\ket{H}$ and $\ket{1}\equiv\ket{V}$. The process matrix with the matrix elements, $\chi^\delta_{t,mn}$, which describes the dynamical mapping of the two qubits, is derived as
\begin{equation}\label{2.X}
\chi^{\delta}_{t}=\frac{1}{4}
\left[
\begin{matrix}
A & Z_{1} & B\\
Z^{\dagger}_{1} & Z_{2} & Z^{\dagger}_{1}\\
B^{\dagger} & Z_{1} & A\\
\end{matrix}
\right],
\end{equation}
where $A$ and $B$ are $6\times6$ matrices with the forms
\begin{equation}
A_{kj}\!\!=\!\!\left\{
\!\!\begin{array}{rcl}
1 & &\!\! k=j=1,6\\
\kappa_{2}(t) & &\!\! k=1, j=6\\
\kappa^{\ast}_{2}(t) & &\!\! k=6, j=1\\
0 & &\!\! \rm{otherwise}
\end{array} \right.,
B_{kj}\!\!=\!\!\left\{
\!\!\begin{array}{rcl}
\kappa_{1}(t) & &\!\! k=j=1,6\\
\kappa_{12}(t) & &\!\! k=1, j=6\\
\Lambda_{12}(t) & &\!\! k=6, j=1\\
0 & &\!\! \rm{otherwise}
\end{array} \right., \nonumber
\end{equation}
in which $Z_{1}$ is a $6\times4$ zero matrix, and $Z_{2}$ is a $4\times4$ zero matrix. Here, $\kappa_{1}(t)$, $\kappa_{2}(t)$, $\kappa_{12}(t)$ and $\Lambda_{12}(t)$ denote the decoherence functions of the dephasing model  \cite{laine2012}. In terms of the Fourier transform of the joint frequency distribution, $P(\omega_1,\omega_2)\equiv|g(\omega_{1},\omega_{2})|^2$ \cite{laine2012}:
\begin{eqnarray}\label{2.k}
G(\tau_{1}, \tau_{2})= &&\int d\omega_{1}d\omega_{2}P(\omega_{1}, \omega_{2})e^{-i\Delta n(\omega_{1}\tau_{1}+\omega_{2}\tau_{2})} \nonumber \\
= && e^{\frac{1}{2}[i\omega_{0} \Delta n(\tau_{1}+\tau_{2})-C\Delta n^{2}(\tau_{1}^{2}+\tau_{2}^{2}+2K\tau_{1}\tau_{2})]},
\end{eqnarray}
the decoherence functions can be expressed as $\kappa_{1}(t)=G(\tau_{1},0)$, $\kappa_{2}(t)=G(0, \tau_{2})$, $\kappa_{12}(t)=G(\tau_{1}, \tau_{2})$, $\Lambda_{12}(t)=G(\tau_{1}, -\tau_{2})$, where $\omega_{0}$ is the center frequency of the pump pulse; $\tau_{1}$ and $\tau_{2}$ are the active times of the quartz plate in spatial mode 1 and spatial mode 2, respectively; and $C$ is the frequency variance of the single photon generated by the SPDC process.  

The local dynamics of the subsystems can also be described by a process matrix, denoted as $\chi'_{t}$, which maps the input space of the different subsystems to the same output space, i.e.,
\begin{equation}\label{2.5}
\rho^{Si}_t=\chi'_{t}(\rho^{Si}_0),
\end{equation}
with the process matrix
\begin{equation}\label{2.6}
\chi'_{t}=\frac{1}{2}
\left[
\begin{matrix}
1 & 0 & 0 & \kappa_{1}(t)\\
0 & 0 & 0 & 0\\
0 & 0 & 0 & 0\\
\kappa^{\ast}_{1}(t) & 0 & 0 & 1\\
\end{matrix}
\right].
\end{equation}

\subsubsection{Reliable process matrices}\label{3.2.3}

Reference \cite{WangKH} and Appendix~\ref{appendix_exp} systematically determined the information of the decoherence functions in Eqs.~(\ref{1.k}, \ref{2.k}) and constructed the process matrix $\chi_{t_{2}}=\chi^{\theta}_{t_{2}}$ [Eq.~(\ref{1.X})] and $\chi_{t_{2}}=\chi^{\delta}_{t_{2}}$ [Eq.~(\ref{2.X})] for the single-photon and two-photon dynamics cases, respectively, by fitting the experimental data reported in Refs.~\cite{liu2011} and \cite{liu2013}.  It was found that the simulated trace distance results obtained from the constructed process matrices were highly consistent with the experimental results in Refs.~\cite{liu2011} and \cite{liu2013}. Hence, the process matrices are considered to be sufficiently reliable to demonstrate our identification of non-Markovianity in the single-photon \cite{liu2011} and two-photon \cite{liu2013} systems based on the concept of QPCs.

\subsection{Non-Markovianity identification by quantum process capabilities of non-CP processes}\label{3.3.2}

\begin{figure}[t]
\includegraphics[width=0.44\textwidth]{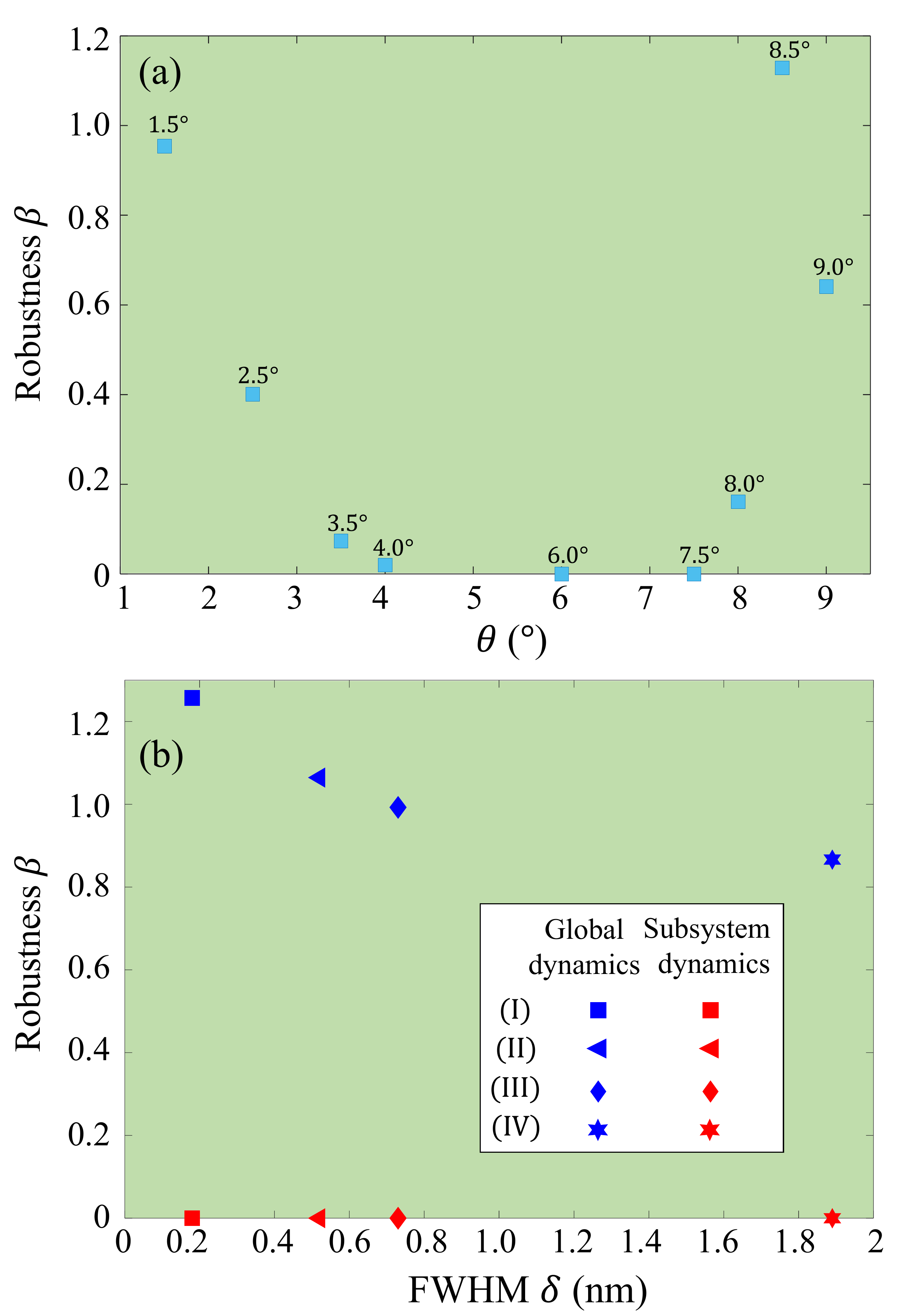}
\caption{Identification results obtained using the criterion (\ref{identification}) of robustness $\beta$ of QPC for non-CP processes in (a) single-photon and (b) two-photon dynamics. We chose to divide the original process $\chi_{t_{2}}$ in half at time $t_{1}$, i.e., $t_{1}=t_{2}/2$. (a) $\beta$ for the environmental conditions created by nine different angles of the cavity in the single-photon system. (b) $\beta$ for global dynamics (blue) and subsystem dynamics (red) for four frequency correlation conditions (\ref{conditions}) in the two-photon system.}
\label{beta_result}
\end{figure}

In the present study, we chose to divide the original process $\chi_{t_{2}}$ in half at time $t_{1}$ according to Eq.~(\ref{divisibility}), i.e., $t_{1}=t_{2}/2$. An analysis of $\Lambda_{t_{2},t_{1}}$ was then conducted using the robustness of QPC $\beta$ [Eq.~(\ref{beta})] in order to identify the non-Markovianity of $\chi_{t_{2}}$ by using Eq.~(\ref{identification}). The identification results for the single-photon and two-photon dynamics cases are presented in the following:\\

\subsubsection{Single-photon dynamics}\label{3.3.3.1}

We quantified the non-CP process capability by using the criterion of $\beta$ [Eq.~(\ref{identification})] to identify non-Markovianity under the nine environmental conditions, as shown in Fig.~\ref{beta_result}(a). For angles of $\theta \leq 4.0^{\circ}$ and $\theta \geq 8.0^{\circ}$, $\beta$ has a value greater than zero, which confirms the amount of additional operations $\chi'$ in (\ref{beta}) required. By contrast, for $\theta=6.0^{\circ}$ and $7.5^{\circ}$, the photon dynamics $\chi^{\theta}_{t_{2}}$ (\ref{1.X}) has no characteristic of the non-Markovianity, thus no additional operations $\chi'$ are required and $\beta$=0.

Our identification results are consistent with the experimental observation reported in Ref.~\cite{liu2011}. The optimized two initial system states of the single photon were prepared as $\rho^{S}_{0,1}=\ketbra{+}$ and $\rho^{S}_{0,2}=\ketbra{-}$, respectively, where $\ket{\pm}=(\ket{H}\pm\ket{V})/\sqrt{2}$. They showed that for environmental conditions (FP cavity angles) of $\theta = 6.0^{\circ}$ and $7.5^{\circ}$, the photon dynamics could not be determined as non-Markovian dynamics. For angles of $\theta \leq 4.0^{\circ}$ and $\theta \geq 8.0^{\circ}$, the BLP criterion identified non-Markovian photon dynamics. 

\subsubsection{Two-photon dynamics}\label{3.3.3.2}

According to the experimental scenario considered in Ref.~\cite{liu2013}, we examine the non-Markovianity under the following conditions of four different pulse spectra [full width at half maximum (FWHM), denoted as $\delta$] and frequency correlations ($K$), i.e.,
\begin{eqnarray}\label{conditions}
&&\rm  (\uppercase\expandafter{\romannumeral 1})\  \delta =0.18~nm: {\it K}=-0.92,\nonumber\\
&&\rm  (\uppercase\expandafter{\romannumeral 2})\  \delta =0.52~nm: {\it K}=-0.66, \nonumber\\
&&\rm  (\uppercase\expandafter{\romannumeral 3})\  \delta =0.73~nm: {\it K}=-0.55, \\
&&\rm  (\uppercase\expandafter{\romannumeral 4})\ \delta =1.89~nm: {\it K}=-0.17.\nonumber
\end{eqnarray}
By the criterion~(\ref{identification}) and the $\beta$ robustness of QPC results in Fig.~\ref{beta_result}(b), we confirm and quantify the degree of non-CP process of $\Lambda_{t_{2},t_{1}}$ for identifying the non-Markovianity of the global photon dynamics $\chi^{\delta}_{t}$ (\ref{2.X}), while the subsystem photon dynamics $\chi'_{t}$ (\ref{2.6}) has no characteristic of non-Markovianity, $\beta$=0. The strength of non-CP characteristic of QPC and non-Markovianity of the principal system dynamics increased as $|K|$ increased. Thus, it can be inferred that the initial correlations between the local parts of different environments can lead to nonlocal memory effects in the global dynamics~\cite{laine2012}. Therefore our results and the conclusions in Ref.~\cite{liu2013} are the same.

The optimized two initial system states reported in Ref.~\cite{liu2013} were prepared as $\rho^{S}_{0,1}=\ketbra{\phi^{+}}$ and $\rho^{S}_{0,2}=\ketbra{\phi^{-}}$, respectively, where $\ket{\phi^{\pm}}=(\ket{HH}\pm\ket{VV})/\sqrt{2}$. The resulting amounts of non-Markovianity were found to be: $(\rm \uppercase\expandafter{\romannumeral 1})~ \mathcal{N}_{\rm BLP}=0.48$, $(\rm \uppercase\expandafter{\romannumeral 2})~ \mathcal{N}_{\rm BLP}=0.23$, $(\rm \uppercase\expandafter{\romannumeral 3})~ \mathcal{N}_{\rm BLP}=0.14$, $(\rm \uppercase\expandafter{\romannumeral 4})~ \mathcal{N}_{\rm BLP}=0.02$. According to the BLP criterion, the dynamics of the entangled photon pairs was identified as non-Markovian under all four conditions (I)-(IV). The dynamics of the individual photons in the two-photon system can not be classified as non-Markovian.

\section{Witnessing non-CP characteristic and non-Markovianity}\label{4}
This section commences by examining the number of experimental settings required when using the criterion (\ref{identification}) of QPC robustness $\beta$ [Eq.~(\ref{beta})] to quantify the degree of non-CP process for non-Markovianity. A more efficient method based on QPC of non-CP process is then proposed to witness non-CP processes for non-Markovian dynamics in Refs.~\cite{liu2011} and \cite{liu2013}.

\begin{figure}[t]
\includegraphics[width=0.38\textwidth]{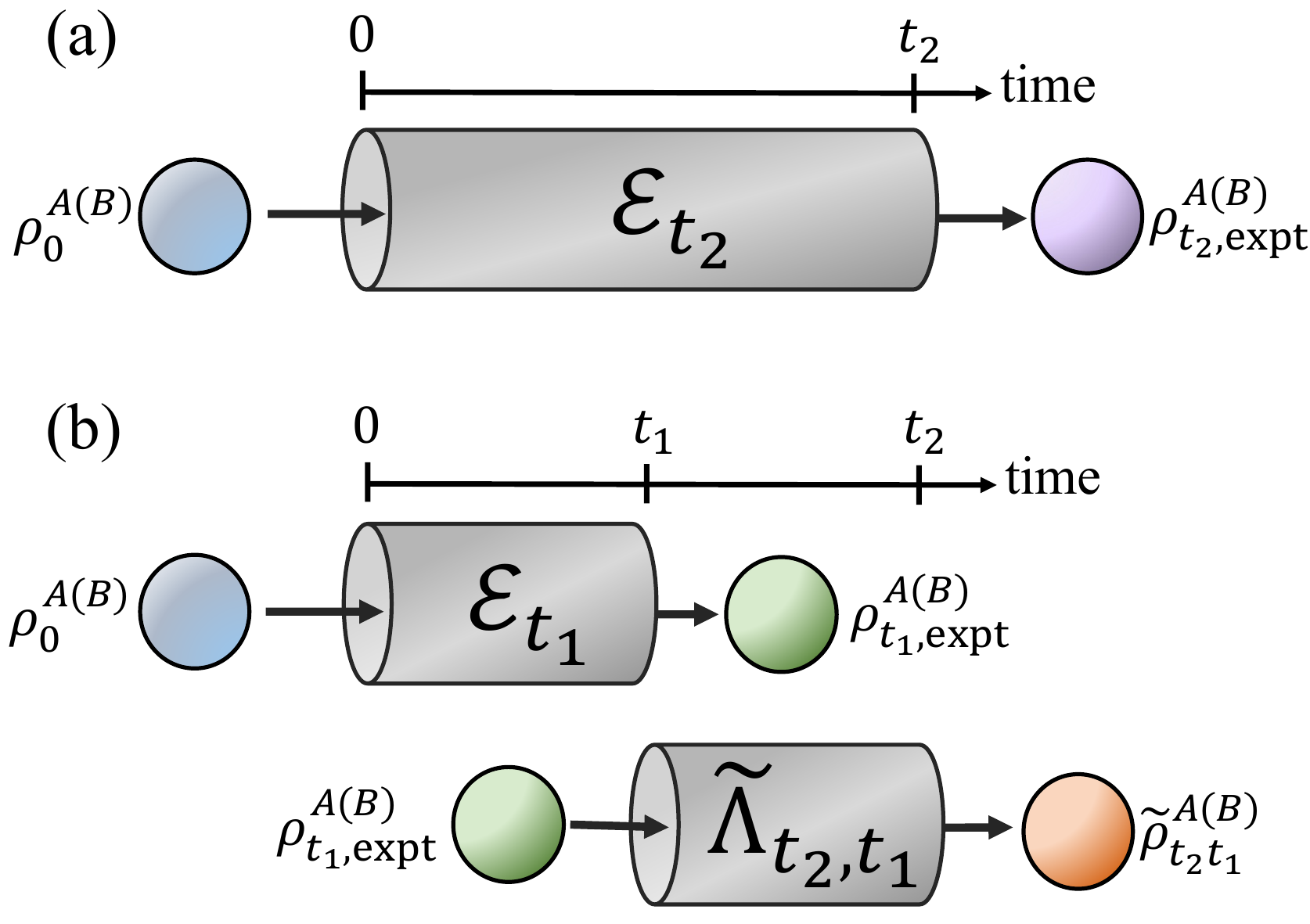}
\caption{Schematic illustration of our witness of non-CP divisibility for non-Markovian dynamics. The underlying idea of the introduced witness~(\ref{Emin2}) is based on violating the composition law (\ref{broken_composition}) by examining the respective output states: (a) $\rho^{A(B)}_{t_2,\mathrm{expt}}=\mathcal{E}_{t_2}(\rho^{A(B)}_0)$, and (b) $\tilde{\rho}^{A(B)}_{t_2,t_1}=\widetilde{\Lambda}_{t_2,t_1}\circ\mathcal{E}_{t_1}(\rho^{A(B)}_0)$.}
\label{Efficient}
\end{figure}

\subsection{Number of measurement settings required for testing the criterion (\ref{identification})}\label{3.4.1}
QPT provides the means to obtain all the information about a process. However, the complete QPT procedure requires the input of several states and the subsequent performance of quantum state tomography (QST). For example, the robustness $\beta$ [Eq.~(\ref{beta})] requires QPT for $\chi_{t_{2}}$ and $\chi_{t_{1}}$ to quantify the value of $\beta$ for the intermediate process $\Lambda_{t_{2},t_{1}}$ though Eq.~(\ref{Determining}).

In single-photon dynamics, a process requires a minimum of four input states to perform QPT to characterize and then construct the process. Furthermore, these four states all require the performance of QST to reconstruct the state by performing experimental measurements of the Pauli observables $X$, $Y$, and $Z$. Thus, a total of twelve experimental settings are required to be performed to complete the QPT procedure.

Similarly, performing QPT for two-photon dynamics requires at least sixteen input states to construct the process. These sixteen states each require the performance of QST to reconstruct the state by performing experimental measurements of the observables, which are products of the Pauli matrices, i.e., $XX$, $XY$, $XZ$, $YX$, $YY$, $YZ$, $ZX$, $ZY$, and $ZZ$. In other words, a total of 144 measurement settings are required to complete the QPT procedure. As described above, when testing the criterion~(\ref{identification}), it is necessary to perform QPT for $\chi_{t_{2}}$ and $\chi_{t_{1}}$. Therefore, a total of 24 and 288 experimental settings are required to characterize single-photon and two-photon dynamics, respectively.

\subsection{Witnessing non-CP divisibility for non-Markovian dynamics}\label{3.4.2}

The basic idea underlying the proposed witness lies in the broken of the composition law~(\ref{CP-divisibility}).
Based on the non-CP divisibility of a non-Markovian dynamics $\mathcal{E}_t$, to guarantee the validity of the composition law~(\ref{CP-divisibility}) for all $0\leq t_1\leq t_2$,
the intermediate process $\Lambda_{t_2,t_1}$ calculated according to Eq.~(\ref{Determining}) should be non-CP for certain $0\leq t_1\leq t_2$.
In other words, if we deliberately replace the intermediate process $\Lambda_{t_2,t_1}$ with a CP one, denoted by $\widetilde{\Lambda}_{t_2,t_1}$, then the composition law will be broken, i.e.,
\begin{equation}\label{broken_composition}
\mathcal{E}_{t_2}\neq\widetilde{\Lambda}_{t_2,t_1}\circ\mathcal{E}_{t_1},
\end{equation}
for certain $0\leq t_1\leq t_2$.

As schematically illustrated in Fig.~\ref{Efficient}, to implement the breaking~(\ref{broken_composition}) for a completely unknown process $\mathcal{E}_t$,
we simply prepare a minimum of two initial states $\rho^{A}_0$ and $\rho^{B}_0$.
Then the non-CP of the original $\Lambda_{t_2,t_1}$, which is also unknown to us, can be witnessed if we are unable to simultaneously equalize $\rho^{A(B)}_{t_2,\mathrm{expt}}=\mathcal{E}_{t_2}(\rho^{A(B)}_0)$ [Fig.~\ref{Efficient}(a)] with $\tilde{\rho}^{A(B)}_{t_2,t_1}=\widetilde{\Lambda}_{t_2,t_1}\circ\mathcal{E}_{t_1}(\rho^{A(B)}_0)$ [Fig.~\ref{Efficient}(b)] with an optimal CP replacement $\widetilde{\Lambda}_{t_2,t_1}$.

Underpinned by this observation, we propose the efficiently computational witness of non-CP $\Lambda_{t_2,t_1}$
for identifying the unknown dynamics $\mathcal{E}_{t_{2}}$ as being non-Markovian:
\begin{equation}\label{Emin2}
\mathcal{W}(\rho^{A}_{0}, \rho^{B}_{0})\equiv \min_{\widetilde{\Lambda}_{t_2,t_1}\geq 0}[\,
\text{tr}(\tilde{\rho}^{A}_{t_2,t_1})
+\text{tr}(\tilde{\rho}^{B}_{t_2,t_1})\,]>2,
\end{equation}
constrained simultaneously by
\begin{equation}
\tilde{\rho}^{A(B)}_{t_2,t_1}-\rho^{A(B)}_{t_2,\mathrm{expt}}\geq 0.
\end{equation}
Consequently, non-Markovianity is identified simply by determining whether the minimum value of the witness kernel $\mathcal{W}(\rho^{A}_{0}, \rho^{B}_{0})$ exceeds 2.
The witness kernel $\mathcal{W}(\rho^{A}_{0}, \rho^{B}_{0})$ can be calculatied by using SDP \cite{sdp1, sdp2} under the constraints $\widetilde{\Lambda}_{t_2,t_1}\geq0$, $\tilde{\rho}^{A(B)}_{t_2,t_1}\geq0$,
and $\tilde{\rho}^{A(B)}_{t_2,t_1}-\rho^{A(B)}_{t_2,\mathrm{expt}}\geq 0$.

When implementing the witness~(\ref{Emin2}), a minimum of merely two initial states $\rho^{A(B)}_{0}$ are prepared experimentally, and the four final states
$\rho^{A(B)}_{t_2,\text{expt}}$ and $\rho^{A(B)}_{t_1,\text{expt}}$ of $\mathcal{E}_{t_2}$ $\mathcal{E}_{t_1}$, respectively, are measured via QST, as shown in Fig.~\ref{Efficient}.
For the case of single-photon dynamics [Fig.~\ref{exp}(a)], these four final states each require the realization of QST, and consequently a total of 12 experimentally measurable settings are required. Here, since $\{\left|H\right\rangle,\left|V\right\rangle,\left|+\right\rangle,\left|R\right\rangle\}$, where $\ket{R}=(\ket{H}+i\ket{V})/\sqrt{2}$, forms an information-complete basis for determining a quantum state, each QST requires only 3 measurement settings. Therefore, it is much smaller than the required measurement settings in performing process tomography for $\beta$ (\ref{beta}) [see also Eq.~(\ref{identification})].
Similarly, for the case of two-photon dynamics [Fig.~\ref{exp}(b)], the four final states each require the implementation of QST, and hence a total of 36 measurement settings are required.
Thus, when using the witness for identifying non-Markovian dynamics, the total number of measurement settings is reduced by 12 compared to the case when using the criterion~(\ref{identification}) of robustness $\beta$ [Eq.~(\ref{beta})] for single-photon dynamics and by 252 for the case of two-photon dynamics.
However, it is worth noting here that the identification results can be affected by the chosen initial states. Therefore, it may be necessary to perform more than one trial.\\

\begin{table*}[t]
\caption{Witnessing single-photon non-Markovian dynamics with two initial states $\rho^{A}_{0}$, $\rho^{B}_{0}$ via the criterion (\ref{Emin2}).}
\begin{ruledtabular}
\begin{tabular}{cccccccccc}
\multicolumn{1}{c}{\text{~~Seleted states~~}}  & \multicolumn{9}{c}{\text{Conditions ($\theta$)}}                                                                                                                                                                                              \\ \hline 
\multicolumn{1}{c}{$\rho^{A}_{0}$, $\rho^{B}_{0}$}             & \text{~~$1.5^{\circ}$}              & \text{$2.5^{\circ}$}              & \text{$3.5^{\circ}$}              & \text{$4.0^{\circ}$}              & ~~~~~\text{$6.0^{\circ}$}~~~~~& ~~~~~\text{$7.5^{\circ}$}~~~~~ & \text{$8.0^{\circ}$}              & \text{$8.5^{\circ}$}              & \text{$9.0^{\circ}$}              \\ \hline
\multicolumn{1}{c}{$\ketbra{H}$, $\ketbra{V}$} & ~~2                           & 2                              & 2                           & 2                             & 2                                   & 2                              & 2                                & 2                              & 2                           \\
\multicolumn{1}{c}{$\ketbra{H}$, $\ketbra{+}$} & ~~2.2325                    & 2.1752                    & 2.0577                    & 2.0179                 & 2                                   & 2                              & 2.1078                    & 2.2464                    & 2.2125                    \\
\multicolumn{1}{c}{$\ketbra{H}$, $\ketbra{-}$} & ~~2.2325                    & 2.1752                    & 2.0577                    & 2.0179                  & 2                                   & 2                              & 2.1078                    & 2.2464                    & 2.2125                    \\
\multicolumn{1}{c}{$\ketbra{H}$, $\ketbra{R}$} & ~~2.2325                    & 2.1752                    & 2.0577                    & 2.0179                & 2                                   & 2                              & 2.1078                    & 2.2464                    & 2.2125                    \\
\multicolumn{1}{c}{$\ketbra{H}$, $\ketbra{L}$} & ~~2.2325                    & 2.1752                    & 2.0577                    & 2.0179                & 2                                   & 2                               & 2.1078                    & 2.2464                    & 2.2125                    \\
\multicolumn{1}{c}{$\ketbra{V}$, $\ketbra{+}$} & ~~2.2325                    & 2.1752                    & 2.0577                    & 2.0179                & 2                                   & 2                              & 2.1078                    & 2.2464                    & 2.2125                    \\
\multicolumn{1}{c}{$\ketbra{V}$, $\ketbra{-}$} & ~~2.2325                    & 2.1752                    & 2.0577                    & 2.0179                 & 2                                  & 2                              & 2.1078                    & 2.2464                    & 2.2125                    \\
\multicolumn{1}{c}{$\ketbra{V}$, $\ketbra{R}$} & ~~2.2325                    & 2.1752                    & 2.0577                    & 2.0179                & 2                                 & 2                               & 2.1078                    & 2.2464                    & 2.2125                    \\
\multicolumn{1}{c}{$\ketbra{V}$, $\ketbra{L}$} & ~~2.2325                    & 2.1752                    & 2.0577                    & 2.0179                & 2                                 & 2                               & 2.1078                    & 2.2464                    & 2.2125                    \\
\multicolumn{1}{c}{$\ketbra{+}$, $\ketbra{-}$} & ~~2.9080                    & 2.5342                    & 2.1312                    & 2.0381                 & 2                                 & 2                              & 2.2659                    & 2.9989                    & 2.7328                    \\
\multicolumn{1}{c}{$\ketbra{+}$, $\ketbra{R}$} & ~~2.733                     & 2.4651                    & 2.1270                    & 2.0377                & 2                                & 2                              & 2.2496                    & 2.7974                    & 2.6134                    \\
\multicolumn{1}{c}{$\ketbra{+}$, $\ketbra{L}$} & ~~2.733                     & 2.4651                    & 2.1270                    & 2.0377                & 2                                & 2                               & 2.2496                    & 2.7974                    & 2.6134                    \\
\multicolumn{1}{c}{$\ketbra{-}$, $\ketbra{R}$} & ~~2.733                     & 2.4651                    & 2.1270                    & 2.0377                 & 2                                & 2                               & 2.2496                    & 2.7974                    & 2.6134                    \\
\multicolumn{1}{c}{$\ketbra{-}$, $\ketbra{L}$} & ~~2.733                     & 2.4651                    & 2.1270                    & 2.0377                 & 2                                & 2                               & 2.2496                    & 2.7974                    & 2.6134                    \\
\multicolumn{1}{c}{$\ketbra{R}$, $\ketbra{L}$} & ~~2.9080                    & 2.5342                    & 2.1312                    & 2.0381               & 2                                & 2                               & 2.2659                    & 2.9989                    & 2.7328                    \\ 
\end{tabular} 
\end{ruledtabular}
\label{E_1qR}
\end{table*}

\begin{table*}[t]
\caption{Witnessing non-Markovianity of two-photon global dynamics with two initial states $\rho^{A}_{0}$, $\rho^{B}_{0}$ via the criterion (\ref{Emin2}).}
\begin{ruledtabular}
\begin{tabular}{ccccc}
\multicolumn{1}{c}{\text{~~Seleted states~~}}  & \multicolumn{4}{c}{\text{Conditions (\ref{conditions})}}       \\ \hline 
\multicolumn{1}{c}{$\rho^{A}_{0}$, $\rho^{B}_{0}$}                      & \text{~~~(I)~~~}              & \text{~~~~~~~(II)~~~~~~~}              & \text{~~~~~(III)~~~~~}              & \text{~~~~~(IV)~~~~~}              \\ \hline
\multicolumn{1}{c}{$\ketbra{\phi^{+}}$, $\ketbra{\phi^{-}}$}            & ~~~3.7245~~~                  & 3.0667                   & 2.8188                      & 2.3567                             \\
\multicolumn{1}{c}{$\ketbra{\phi^{+}}$, $\ketbra{\psi^{+}}$}           & ~~~3.7245~~~                  & 3.0667                   & 2.8188                      & 2.3567                             \\
\multicolumn{1}{c}{$\ketbra{\phi^{+}}$, $\ketbra{\psi^{-}}$}            & ~~~3.7245~~~                  & 3.0667                   & 2.8188                      & 2.3567                             \\
\multicolumn{1}{c}{$\ketbra{\phi^{-}}$, $\ketbra{\psi^{+}}$}            & ~~~3.7245~~~                  & 3.0667                   & 2.8188                      & 2.3567                             \\
\multicolumn{1}{c}{$\ketbra{\phi^{-}}$, $\ketbra{\psi^{-}}$}             & ~~~3.7245~~~                  & 3.0667                   & 2.8188                      & 2.3567                             \\
\multicolumn{1}{c}{$\ketbra{\psi^{+}}$, $\ketbra{\psi^{-}}$}            & ~~~2.0594~~~                  & 2.0874                   & 2.0871                      & 2.1801                             \\
\multicolumn{1}{c}{$\ketbra{S_{1}}$, $\ketbra{S_{2}}$}                 & ~~~2.7437~~~                  & 2.4938                   & 2.3925                      & 2.2629                             \\
\multicolumn{1}{c}{$\ketbra{S_{1}}$, $\ketbra{S_{3}}$}                 & ~~~3.1098~~~                  & 2.7337                   & 2.5815                      & 2.3122                             \\
\multicolumn{1}{c}{$\ketbra{S_{1}}$, $\ketbra{S_{4}}$}                 & ~~~3.1098~~~                  & 2.7337                   & 2.5815                      & 2.3122                             \\
\multicolumn{1}{c}{$\ketbra{S_{2}}$, $\ketbra{S_{3}}$}                 & ~~~3.1098~~~                  & 2.7337                   & 2.5815                      & 2.3122                             \\
\multicolumn{1}{c}{$\ketbra{S_{2}}$, $\ketbra{S_{4}}$}                 & ~~~3.1098~~~                  & 2.7337                   & 2.5815                      & 2.3122                             \\
\multicolumn{1}{c}{$\ketbra{S_{3}}$, $\ketbra{S_{4}}$}                 & ~~~2.7437~~~                  & 2.4938                   & 2.3925                      & 2.2629                             \\
\multicolumn{1}{c}{$\ketbra{H H}$, $\ketbra{V V}$}                        & ~~~2~~~                           & 2                          & 2                              & 2                                      \\
\multicolumn{1}{c}{$\ketbra{H H}$, $\ketbra{H V}$}                        & ~~~2~~~                           & 2                          & 2                              & 2                                      \\
\multicolumn{1}{c}{$\ketbra{H H}$, $\ketbra{H +}$}                         &~~~~2~~~                           & 2                          & 2                              & 2                                      \\ 
\multicolumn{1}{c}{$\ketbra{H H}$, $\ketbra{H R}$}                        & ~~~2~~~                           & 2                           & 2                              & 2                                      \\
\multicolumn{1}{c}{$\ketbra{H H}$, $\ketbra{+ +}$}                         & ~~~2.7946~~~                   & 2.4263                  & 2.3038                      & 2.0558                              \\
\multicolumn{1}{c}{$\ketbra{H H}$, $\ketbra{R R}$}                        & ~~~2.7946~~~                   & 2.4263                  & 2.3038                      & 2.0558                              \\ 
\multicolumn{1}{c}{$\ketbra{+ +}$, $\ketbra{- -}$}                           & ~~~3.5892~~~                   & 2.8525                   & 2.6075                      & 2.1116                               \\
\multicolumn{1}{c}{$\ketbra{+ +}$, $\ketbra{H +}$}                         & ~~~3.5892~~~                   & 2.8525                   & 2.6075                      & 2.1116                               \\
\multicolumn{1}{c}{$\ketbra{+ +}$, $\ketbra{H R}$}                         & ~~~2.7946~~~                   & 2.4263                  & 2.3038                      & 2.0558                              \\  
\multicolumn{1}{c}{$\ketbra{+ +}$, $\ketbra{R R}$}                         & ~~~3.5892~~~                   & 2.8525                   & 2.6075                      & 2.1116                               \\ 
\multicolumn{1}{c}{$\ketbra{- -}$, $\ketbra{H -}$}                            & ~~~3.5892~~~                   & 2.8525                   & 2.6075                      & 2.1116                               \\  
\multicolumn{1}{c}{$\ketbra{R R}$, $\ketbra{L L}$}                        & ~~~3.5892~~~                   & 2.8525                   & 2.6075                      & 2.1116                               \\  
\end{tabular} 
\end{ruledtabular}
\label{E_2qR}
\end{table*}

\subsection{Witnessing photonic non-Markovian dynamics}\label{3.4.3}
The experimental feasibility of the proposed witness for identifying non-Markovian dynamics is demonstrated using the reported one-photon \cite{liu2011} and two-photon dynamics \cite{liu2013}. The corresponding results are presented in the following:\\

\subsubsection{Single-photon dynamics}\label{3.4.3.1}

As shown in Table \ref{E_1qR}, nine different environmental conditions were created by rotating the FP cavity [Fig.~\ref{exp}(a)]. Moreover, two different initial states were chosen in each trial. For initial states of $\rho^{A}_{0}$ as $\ketbra{H}$ and $\rho^{B}_{0}$ as $\ketbra{V}$, the witness~(\ref{Emin2}) was found to have a value of $\mathcal{W}=2$ for all environmental conditions, indicating that there existed an intermediate CP process $\Lambda'_{t_{2}t_{1}}=\Lambda_{t_{2},t_{1}}$ in the witness kernel that satisfied both $\rho^{A'}_{t_{2}t_{1}}=\rho^{A}_{t_{2}t_{1}}= \rho^{A}_{t_{2},\text{expt}}$ and $\rho^{B'}_{t_{2}t_{1}}=\rho^{B}_{t_{2}t_{1}}= \rho^{B}_{t_{2},\text{expt}}$.
Given these initial states, it was thus impossible to confirm whether or not the dynamics $\mathcal{E}_{t_{2}}$ is non-Markovian. 
However, when the selected states were chosen as different base states, such as $\ketbra{H}, \ketbra{+}$ and $\ketbra{V}, \ketbra{R}$ and so on, or the same base states with coherence, such as $\ketbra{+}, \ketbra{-}$ and $\ketbra{R}, \ketbra{L}$, the witness had a value of $\mathcal{W}=2$ for cavity angles of $\theta=6.0^{\circ}$ and $7.5^{\circ}$, but $\mathcal{W} > 2$ for all other angles. Here we define $\ket{L}=(\ket{H}-i\ket{V})/\sqrt{2}$. Thus, while the dynamics $\mathcal{E}_{t_{2}}$ at $\theta=6.0^{\circ}, 7.5^{\circ}$ still could not be confirmed as non-Markovian, those at $\theta \leq 4.0^{\circ}$ and $\geq 8.0^{\circ}$ were identified as non-Markovian.\\

\subsubsection{Two-photon dynamics}\label{3.4.3.2}

For the global dynamics of the two photons in Fig.~\ref{exp}(b), two states were chosen from the entangled states as initial states and two states were chosen from the separable states (see Table \ref{E_2qR}). Considering $\ket{\phi^{+}}$ and $\ket{\phi^{-}}$ of the selected Bell states as an example, the witness kernel was found to have a value of $\mathcal{W} > 2$ for all four conditions (I)-(IV) (\ref{conditions}). Hence, the four dynamics were all identified as non-Markovian. Other selection combinations of the Bell states: $\ket{\phi^{\pm}}$ and $\ket{\psi^{\pm}}=(\ket{HV}\pm\ket{VH})/\sqrt{2}$, and entangled states, e.g., $\ket{S_1}=(\ket{H+}+\ket{V-})/\sqrt{2}$, $\ket{S_2}=(\ket{H+}-\ket{V-})/\sqrt{2}$, $\ket{S_3}=(\ket{H-}+\ket{V+})/\sqrt{2}$, $\ket{S_4}=(\ket{H-}-\ket{V+})/\sqrt{2}$, all yield the same identification results. When selecting separate states, for example, $\ket{H H}$ and $\ket{V V}$, as the initial states, the four dynamics yield $\mathcal{W}=2$ and thus can not be confirmed as either non-Markovian or Markovian. However, all four dynamics were identified as non-Markovian when states with coherence, such as $\ket{+ +}$ and $\ket{- -}$ or $\ket{+ +}$ and $\ket{R R}$, were chosen as the initial states. 

For the local dynamics of these composite dynamics, we take the photons in arm $1$ passing through QP1 as an example, the complete process of QP1 was divided into two subprocesses, and the two initial states were chosen as the same basis states or different basis states, that is, $\ket{H}, \ket{V}$ and $\ket{H}, \ket{+}$, and so on. The witness kernel has a value of $\mathcal{W}=2$ under all four experimental conditions (I)-(IV) (\ref{conditions}). Hence, the local dynamics can not be confirmed as non-Markovian.\\

\subsubsection{Discussion}

A further investigation was performed to evaluate the feasibility of the proposed efficient identification method when preparing only one initial state. For this case, Eq.~(\ref{Emin2}) was reformulated as
\begin{equation}\label{E_min1}
\mathcal{W}(\rho^{A}_{0})\equiv \min_{\tilde{\Lambda}_{t_{2},t_{1}}\geq0}[\,\text{tr}(\tilde{\rho}^{A}_{t_2,t_1}))]>1,
\end{equation}
under the constraints of $\widetilde{\Lambda}_{t_2,t_1}\geq0$, $\tilde{\rho}^{A}_{t_2,t_1}\geq0$,
and $\tilde{\rho}^{A}_{t_2,t_1}-\rho^{A}_{t_2,\mathrm{expt}}\geq 0$. We examine the corresponding identification results for the single-photon and two-photon dynamics, respectively. In both cases, the identification results are unable to confirm the existence of non-Markovian dynamics.

According to these results, witnessing non-Markovian dynamics based on QPC can be performed only when selecting two initial states with quantum characteristics, i.e., superposition and entanglement. In such cases, the identification results are consistent with those obtained using the criterion (\ref{identification}) of robustness $\beta$ [Eq.~(\ref{beta})]. They are all consistent with the experimental results presented in Refs.~\cite{liu2011}~and~\cite{liu2013}.

\section{Conclusion and outlook}\label{5}

To identify and measure non-Markovian dynamics without examining all initial system states and using entangled pairs compared to the BLP \cite{breuer2009,laine2010} and RHP \cite{rivas2010} criteria, we prove and utilize the new quantum process capabilities (QPCs) of non-CP processes. The robustness of a non-CP process (\ref{beta}) is a sensible QPC measure satisfying the properties (MP1), (MP2) and (MP3). This implies that non-Markovianity can also be quantified by QPCs of non-CP processes. For experimental implementations of the introduced identification~(\ref{identification}) and measure~(\ref{betanonMK}) based on QPCs of non-CP processes, they can be systematically realized using quantum process tomography (QPT). To illustrate further the utility of the new QPCs of non-CP processes, we also introduce a witness to identify non-Markovianity using at least four system states without QPT.

To specifically show the feasibility of the introduced concept and method, we demonstrate that, with all-optical setups, the non-Markovianity of single-photon and two-photon dynamics in birefringent crystals can be analyzed and identified. For both photonic dynamics, our identification results (Fig.~\ref{beta_result} and Tables \ref{E_1qR} and \ref{E_2qR}) are faithful and consistent with the experimental observation reported in Refs.~\cite{liu2011,liu2013}. For example, the dynamics of the individual photons in the two-photon system can not be classified as non-Markovian. Still, the global dynamics does, as a signature of the nonlocal memory effects~\cite{laine2012}.

With the benefits underlying QPCs of non-CP processes, our methods proposed in this study can be employed to explore non-Markovianity and the related effects on quantum-information processing in other non-Markovian dynamical systems where quantum tomography of state and process is implementable, such as the estimation error threshold of non-Markovian effect-tolerant quantum computation in superconducting systems \cite{terhal2005fault, aharonov2006fault}.

The above summary and conclusion motivate several questions for future work: Apart from the robustness and witness shown here, do other identification methods exist for non-Markovianity? If this is the case, can they satisfy all the conditions for a sensible measure of QPC? Moreover, in addition to the relationship between non-Markovianity and QPC of non-CP processes, can channel resource theory~\cite{Takagi19} relate to QPC and also describe and examine non-Markovianity for practical experiments? Finally, because quantum state tomography (QST) is required for the proposed methods, how estimations of final state information without competing for QST can aid more efficient non-Markovianity identification may be worth further investigation.

\acknowledgements

C.-M.L. acknowledges the partial support from the National Science and Technology Council, Taiwan, under Grant No.~111-2112-M-006-033, No.~111-2119-M-007-007, No.~111-2123-M-006-001, and No.~107-2628-M-006-001-MY4. H.-B.C. is partially supported by the National Science and Technology Council, Taiwan, under Grant No. NSTC 111-2112-M-006-015-MY3. This work is also supported partially by the National Center for Theoretical Sciences, Taiwan.

\appendix

\section{Breuer-Laine-Piilo (BLP) and Rivas-Huelga-Plenio (RHP) criteria}\label{appendix_rev}

\subsection{BLP criterion for non-Markovianity}\label{3.1.2}

The BLP criterion~\cite{breuer2009,laine2010} focuses on the time evolution of the trace distance $D(\rho_{1,2};t)=\|\mathcal{E}_t(\rho_1)-\mathcal{E}_t(\rho_2)\|_1/2$ of two arbitrary initial system states, where $\|A\|_1=\tr\sqrt{A^\dagger A}$ denotes the trace norm of a matrix $A$. One of the critical properties of the trace distance is the contraction under a positive map. Therefore, for a Markovian dynamics $\mathcal{E}_t$ satisfying the composition law~(\ref{CP-divisibility}), the CP-divisibility guarantees that the trace distance always decreases monotonically in time. On the contrary, any revival of the trace distance is a confident evidence showing the violation of the CP-divisibility, witnessing the non-Markovian nature of the dynamics.

\subsubsection{Identifying non-Markovianity}

When considering non-Markovianity identification instead a non-Markovianity measure of the total dynamics, the signature that any revival of the trace distance can attain the aim of identifying non-Markovianity. However, such identification still requires maximization over all possible initial state pairs $(\rho_1,\rho_2)$ to get the best revival visibility. 

\subsubsection{Measuring non-Markovianity}

Ultimately, the BLP criterion help define the following non-Markovianity measure which quantifies the non-Markovianity according the maximal revival of the trace distance
\begin{equation}
\mathcal{N}_\mathrm{BLP}= \max_{(\rho_1,\rho_2)}\int_{\frac{\partial D}{\partial t}>0}
\left[\frac{\partial D(\rho_{1,2};t)}{ \partial t}\right]dt.
\end{equation}
This measure requires a maximization over all possible initial state pairs $(\rho_1,\rho_2)$.

From an information-theoretic viewpoint, it determines the probability $[D(\rho_{1,2};t)+1]/2$ of successfully guessing the correct states. In this sense, the trace distance encodes our knowledge about a system and the contraction of trace distance under a positive map implies the destruction of the knowledge, reflecting the so-called information processing theory. Consequently, the BLP theory explains the monotonical decreasing of trace distance as a flow of information out of the system; on the other hand, the revival of trace distance is conceived as a backflow of information from the environment to the system.

\subsection{RHP criterion for non-Markovianity}\label{3.1.3}

From the discussions in Sec.~\ref{3.1.1}, it can be understood that the origin of non-Markovinaity can be mathematically expressed as the violation of the CP-divisiblity. Therefore, in the RHP criterion \cite{rivas2010}, non-Markovianity is quantified by the degree of deviation from being CP-divisible. According to Eq.~(\ref{CP-divisibility}), a dynamics is CP-divisible if $\Lambda_{t_2,t_1}$ is CP for all $0\leq t_1\leq t_2$. Meanwhile, $\Lambda_{t_2,t_1}$ is CP if and only if $\| (\Lambda_{t_2,t_1}\otimes\mathbb{I}^{A})\ket{\Psi^{SA}}\!\!\bra{\Psi^{SA}} \|_1=1$, where $\ket{\Psi^{SA}}$ is a maximally entangled state between the system and a copy of well-isolated ancilla possessing the same degrees of freedom of the system.

\subsubsection{Identifying non-Markovianity}

According to the above discussion, on the other hand, if $\Lambda_{t_2,t_1}$ is non-CP, then we have $\| (\Lambda_{t_2,t_1}\otimes\mathbb{I}^{A})\ket{\Psi^{SA}}\!\!\bra{\Psi^{SA}} \|_1>1$. Therefore, this feature of non-Markovian dynamics
can be used to identify non-Markovianity. It is clear that implementing this identification requires an entanglement source of $\ket{\Psi^{SA}}$ and the entanglement-related manipulation such as the well isolation of ancilla.

\subsubsection{Measuring non-Markovianity}

The RHP criterion underlying the above non-Markovianity identification can be used to define the degree of non-Markovianity of the overall dynamics by detecting the CP-divisibility in an infinitesimal time interval as
\begin{equation}\label{RHPnonMK}
\mathcal{N}_{\rm RHP} =\int^{\infty}_{0} \lim\limits_{\delta' \rightarrow 0^{+}} \frac{\| (\Lambda_{t+\delta’,t}\otimes\mathbb{I}^{A})\ket{\Psi^{SA}}\!\!\bra{\Psi^{SA}}\|_1 -1}{\delta'}dt.
\end{equation}
The requirements for experimental implementation are the same as the identification reviewed above.

\section{Process matrices of the two-photon system: $\chi^{\delta}_{t}$ (\ref{2.X}) and $\chi'_{t}$ (\ref{2.6})}\label{appendix_exp}

In the following we detail how to utilize the reported experimental data in Ref.~\cite{liu2013} to construct faithful process matrices $\chi^{\delta}_{t}$ (\ref{2.X}) and $\chi'_{t}$ (\ref{2.6}) for demonstrating our methods presented in Secs.~\ref{3.3.2} and~\ref{3.4.3}.

\begin{table*}[t]
\caption{Parameters used in fitting the experimental frequency spectra of the pump pulses and trace distance reported in Ref.~\cite{liu2013}. The simulated frequency distributions and trace distance dynamics using these parameters are close to the results presented in Ref.~\cite{liu2013}. Given the parameters, $K$, $\omega_{0}$, $C$, and $\Delta n$, the $G(\tau_{1}, \tau_{2})$ can be determined and then the process matrices $\chi^{\delta}_{t}$ (\ref{2.X}) and $\chi'_{t}$ (\ref{2.6}) can be faithfully derived from the experimental data reported in Ref.~\cite{liu2013}.}
\begin{ruledtabular}
\begin{tabular}{cccccccc}
\textrm{Conditions [compared to Eq.~(\ref{conditions})]}&
\textrm{$K$}&
\textrm{$\omega_{0}~(\rm nm)$}&
\textrm{$\delta ~(\rm nm)$}&
\textrm{$C$}&
\textrm{$\Delta n$}\\
\colrule
$\rm (\uppercase\expandafter{\romannumeral 1})$ &  $-0.9174$ & $389.7235$  & $0.1799$ & $0.0233$ & $0.0444$\\
$\rm (\uppercase\expandafter{\romannumeral 2})$ &  $-0.6655$ & $389.7692$  & $0.5181$ & $0.1950$ & $0.0156$\\
$\rm (\uppercase\expandafter{\romannumeral 3})$ &  $-0.5564$ & $389.7436$  & $0.7305$ & $0.3844$ & $0.0115$\\
$\rm (\uppercase\expandafter{\romannumeral 4})$ &  $-0.1645$ & $390.0128$  & $1.8885$ & $2.5760$ & $0.0045$\\
\end{tabular}
\end{ruledtabular}
\label{allparameters}
\end{table*}

\begin{figure*}[t]
\includegraphics[width=18cm]{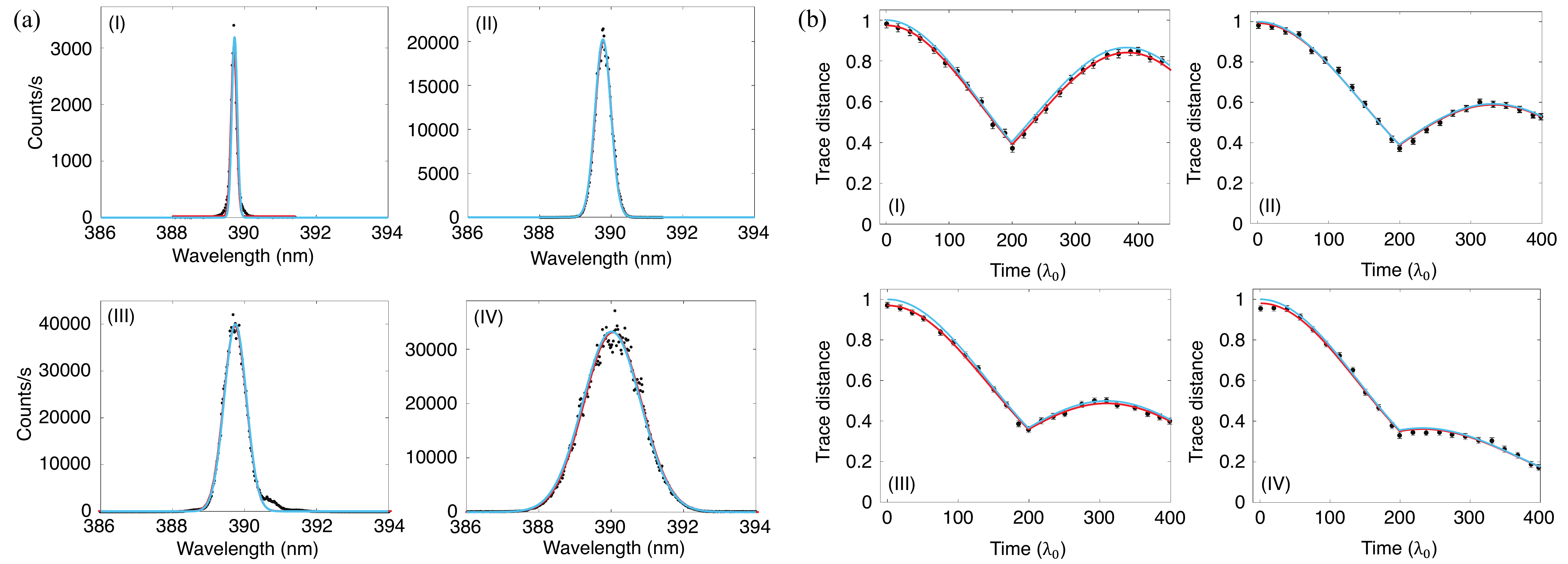}
\caption{Simulated results for faithful process matrix constructions. (a) Simulated frequency spectra of pump pulses. The simulation results of spectra (blue lines) are obtained by using the Gaussian function to fit the experimental data (black dots and red lines) reported in Ref.~\cite{liu2013}. The count rate also corresponds to Fig.~5 in Ref.~\cite{liu2013}. The fitting parameters are shown in Table~\ref{allparameters}. (b) Simulations of trace distance under the
four frequency correlation conditions (I)-(IV) (\ref{conditions}). The $\lambda_{0}$=780 nm denotes the effective path difference corresponding to Ref.~\cite{liu2013}. We utilize the parameters listed in Table.~\ref{allparameters} and substitute into Eq.~(\ref{TD}) to simulate the trace distance of the two-photon
polarization states (blue lines). The simulated trace distance dynamics are highly consistent with
the experimental trace distance dynamics (black dots and red lines) illustrated in Ref.~\cite{liu2013}. The evaluated amounts of non-Markovianity close to the reported ones are: (\rm \uppercase\expandafter{\romannumeral 1}) $\mathcal{N_{\rm BLP}}=0.46$, (\rm \uppercase\expandafter{\romannumeral 2}) $\mathcal{N_{\rm BLP}}=0.20$, (\rm \uppercase\expandafter{\romannumeral 3}) $\mathcal{N_{\rm BLP}}=0.13$, and (\rm \uppercase\expandafter{\romannumeral 4}) $\mathcal{N_{\rm BLP}}=0.01$. See the summary of the reported quantities of $\mathcal{N_{\rm BLP}}$ at the end of Sec.~\ref{3.3.3.2}.}
\label{expdata}
\end{figure*}

\subsection{Criterion for non-Markovian dynamics}

In order to implement the BLP criterion, Liu \textit{et al.} \cite{liu2013} in their experiment prepared the two initial system states of two photons generated from the SPDC process as $\ket{\phi^{+}}$ and $\ket{\phi^{-}}$. The related details of physical scenarios and interaction between principle system and environments can also be found in the main text, \ref{3.2.1}. The time evolution of trace distance of $\ket{\phi^{+}}$ and $\ket{\phi^{-}}$, is found to be
\begin{equation}\label{TD}
D(t)={\rm exp} \left[-\frac{1}{2} \Delta n^{2}C(\tau_{1}^{2}+\tau_{2}^{2}-2\left|K\right| \tau_{1}\tau_{2})\right].
\end{equation}

\subsection{Simulation method}

To objectively compare our results with the conclusion of Laine \textit{et al.}~\cite{laine2012} and Liu \textit{et al.}~\cite{liu2013}, we systematically exploit the experimental data reported in Ref.~\cite{liu2013} to determine the Fourier transform of the joint frequency distribution, $G(\tau_{1}, \tau_{2})$ (\ref{2.k}), required to construct the process matrices $\chi^{\delta}_{t}$ (\ref{2.X}) and $\chi'_{t}$ (\ref{2.6}). The steps of the simulation method are detailed as follows.

First, we use the following function:
\begin{equation}
f_{1}(x)=\tilde{X} {\rm exp}[-\tilde{Y}x^{2}],\label{f1} 
\end{equation}
to fit the experimental data of trace distance reported in Ref.~\cite{liu2013} when the quartz plate in spatial mode 1 (QP1) is active. Thus, we can find the parameters $\tilde{X}$ and $\tilde{Y}$. The parameter $\tilde{Y}$ is obviously equal to $\Delta n^{2}C/2$ in Eq.~(\ref{TD}).

Second, when the quartz plate in spatial mode 2 (QP2) is active, we use the function:
\begin{eqnarray}
&&f_{2}(x)=\tilde{X} {\rm exp} [-\tilde{Y}(199^{2}+(x-199)^{2}\nonumber\\
&&\ \ \ \ \ \ \ \ \ \ \ \ \ \ \ \ \ \ \ \ -2|K|199(x-199))],\label{f2} 
\end{eqnarray}
to fit the experimental data. Since $\tilde{X}$ and $\tilde{Y}$ are obtained from the first fitting step, the frequency correlation $K$ can be determined by the present fitting step. 

Third, we use the following Gaussian function:
\begin{equation}
f_{3}(x)=\frac{1}{\sqrt{2\pi}\sigma_{0}}\text{exp}\left[-\frac{(x-\omega_{0})^2}{2 \sigma_{0}^{2}}\right],  
\end{equation}
to fit the experimental data of the frequency spectra of pump pulses shown in Ref.~\cite{liu2013}. From this fitting procedure, we can obtain the frequency center, $\omega_{0}$, and standard deviation, $\sigma_{0}$, of pump pulse spectra. According to the Gaussian distribution, we convert standard deviation to FWHM (full width at half maximum) using $\delta=\sigma_{0}\sqrt{8\rm ln 2} $, where $\delta$ is FWHM of pump pulse spectra. The fitting parameters are shown in Table~\ref{allparameters}.
To compare our simulating results with the experimental outcomes, they are depicted together in Fig.~\ref{expdata}(a). As illustrated, the simulated spectra are highly consistent with the results reported in Ref.~\cite{liu2013}.

Fourth, to determine the unknown parameters $\Delta n$ and $C$, we utilize the numerical values in the third step and make an assumption about the frequency spectra of the single photon created by SPDC. That is, we assume that FWHM of the frequency spectrum of the single photon generated by SPDC is two times broader than that of the pump pulse, and the frequency spectrum of the single photon satisfies Gaussian distribution. Consequently, we obtain an estimate of frequency variance of the single photon created by SPDC through $C=(2\delta/\sqrt{8\rm ln 2})^2$, in accordance with the Gaussian distribution. The rest unknown parameter $\Delta n$ can be acquired easily by evaluating $\tilde{Y}=\Delta n^{2}C/2$. 

\subsection{Reliable process matrices $\chi^{\delta}_{t}$ (\ref{2.X}) and $\chi'_{t}$ (\ref{2.6})}

All parameters we obtained through the experimental data in Ref.~\cite{liu2013} are listed in Table.~\ref{allparameters}. They are then used to construct the process matrices $\chi^{\delta}_{t}$ (\ref{2.X}) and $\chi'_{t}$ (\ref{2.6}), and by which the time evolution of trace distance of $\ket{\phi^{+}}$ and $\ket{\phi^{-}}$ can be obtained as well. Finally, we simulate the trace distance under the four different conditions (I)-(IV) [see Eq.~(\ref{conditions})] by the fitting parameters we obtained. The corresponding results are shown in Fig.~\ref{expdata}(b). As shown, our simulations almost reproduce the results reported in Ref.~\cite{liu2013}. A very small mismatch of the quantities of $\mathcal{N_{\rm BLP}}$ between the simulations and the measure reported in Ref.~\cite{liu2013} is due to the error in the fitting procedure. Thus, we conclude that the underlying process matrices $\chi^{\delta}_{t}$ (\ref{2.X}) and $\chi'_{t}$ (\ref{2.6}) are faithfully constructed from the experimental data shown in Ref. \cite{liu2013}.

\end{document}